\documentclass{article}
\usepackage{spconfa4,amsmath,graphicx,subcaption,caption}
\usepackage[usenames,dvipsnames]{color}
\usepackage{siunitx,,amssymb}

\usepackage{tikz}
\usetikzlibrary{shapes,arrows}
\usetikzlibrary{calc}
\usetikzlibrary{positioning}

\usepackage{balance}
\usepackage[sort]{cite}
\usepackage{flushend}

\usepackage[hidelinks]{hyperref}

\usepackage{acronym}
\acrodef{AEC}{acoustic echo cancellation}
\acrodef{SRO}{sample rate offset}
\acrodef{AIR}{acoustic impulse response}
\acrodef{ppm}{parts-per-million}
\acrodef{STFT}{short-time Fourier transform}
\acrodef{FIR}{finite impulse response}
\acrodef{CD}{coherence drift}
\acrodef{DWACD}{dynamic weighted average coherence drift}
\acrodef{GCC}{generalized cross-correlation}
\acrodef{ERLE}{echo return loss enhancement}
\acrodef{MOS}{mean opinion score}
\acrodef{PESQ}{perceptual evaluation of speech quality}
\acrodef{RIR}{room impulse response}

\title{Sample Rate Offset Compensated Acoustic Echo Cancellation \\ for Multi-Device Scenarios}
\name{Srikanth Korse, Oliver Thiergart, and Emanu\"{e}l A.~P.~Habets}
\address{International Audio Laboratories Erlangen$^\dag$\thanks{$^\dag$A joint institution of the Friedrich-Alexander-Universit\"{a}t Erlangen-N\"{u}rnberg (FAU) and Fraunhofer IIS, Germany.}, Am Wolfsmantel 33, 91058 Erlangen, Germany}
\begin{document}
\ninept
\maketitle
\sloppy
\begin{abstract}
\Ac{AEC} in multi-device scenarios is a challenging problem due to \ac{SRO} between devices. The \ac{SRO} hinders the convergence of the \ac{AEC} filter, diminishing its performance. To address this , we approach the multi-device \ac{AEC} scenario as a multi-channel \ac{AEC} problem involving a multi-channel Kalman filter, \ac{SRO} estimation, and resampling of far-end signals. Experiments in a two-device scenario show that our system mitigates the divergence of the multi-channel Kalman filter in the presence of \ac{SRO} for both correlated and uncorrelated playback signals during echo-only and double-talk. Additionally, for devices with correlated playback signals, an independent single-channel \ac{AEC} filter is crucial  to ensure fast convergence of~\ac{SRO} estimation.
\end{abstract}
\begin{keywords}
Acoustic Echo Cancellation, Kalman Filtering, Sample Rate Offset, Multi-Device
\end{keywords}

\acresetall
\vspace{-2mm}
\section{Introduction}
\label{sec:intro}
\begin{figure}[t]
\centering
  \includegraphics[width=0.9\linewidth]{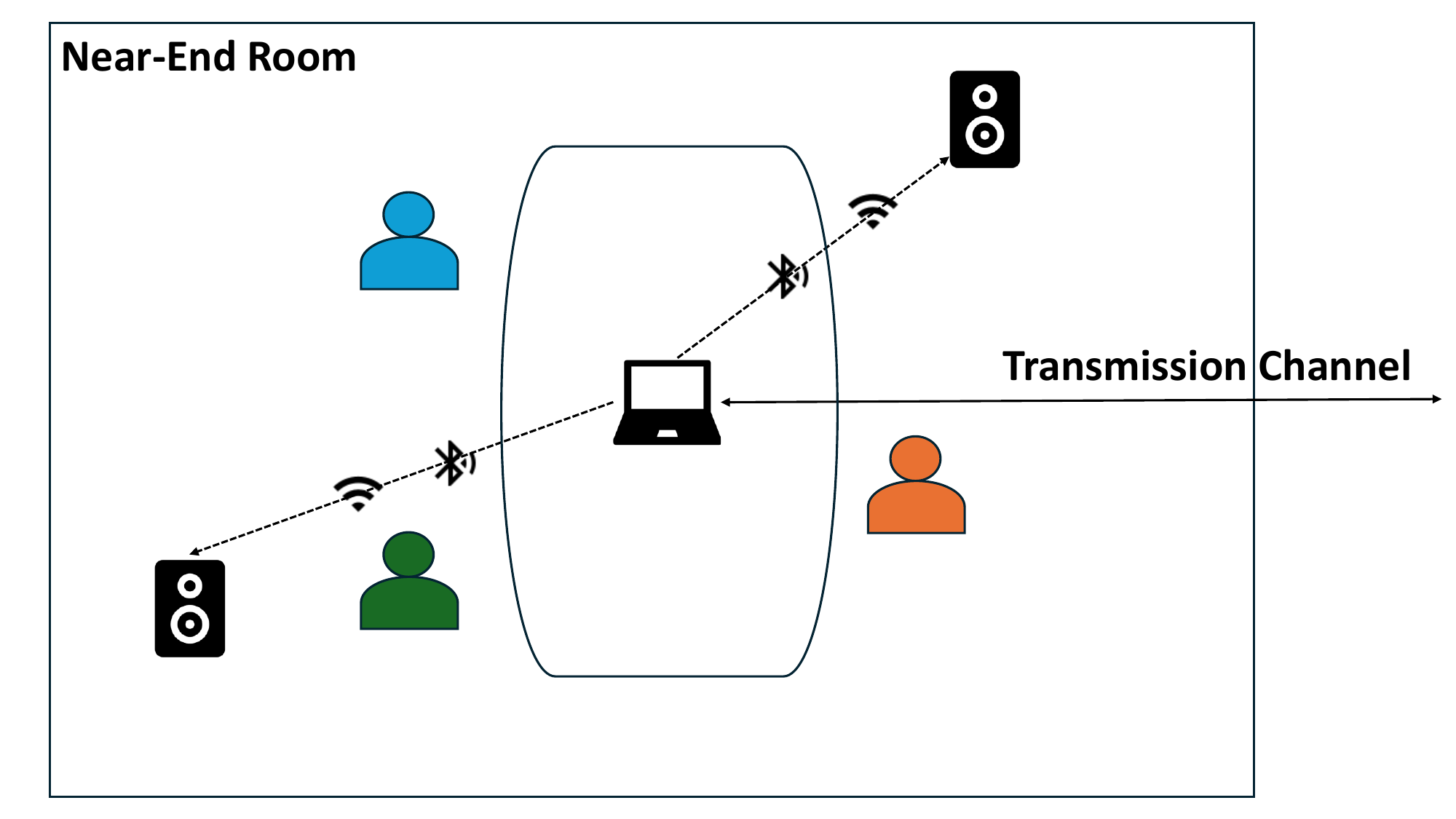}
  \caption{ Spatial teleconferencing system with the help of a main client connected to auxiliary devices via Bluetooth or WiFi.\vspace{-9pt}}
  \label{fig:spatial_teleconferencing_use_case}
\end{figure}

\vspace{-2mm}
\Ac{AEC} is a critical acoustic signal processing technique employed to mitigate the echo caused by the acoustic coupling between loudspeakers and microphones~\cite{Haensler_2004_book,enzner_frequency-domain_2006, malik_recursive_2011, kuech_state-space_2014}. In spatial teleconferencing systems~\cite{herre2012interactive} using multiple devices, as illustrated in Fig.~\ref{fig:spatial_teleconferencing_use_case}, a primary device (such as a smartphone or a laptop) is typically hosts the conferencing application. The host is connected to multiple auxiliary loudspeakers via Bluetooth or WiFi. This setup is susceptible to \ac{SRO}, also known as clock drifts, due to the fact that each device operates on its own individual clock~\cite{guggenberger_clockdrift_2014,he_analysis_2015}. The \ac{AEC} algorithm running on the primary device must eliminate echoes originating from its own loudspeaker and those from the auxiliary loudspeakers in the presence of \ac{SRO}. 
In the presence of \ac{SRO}, the misalignment of the microphone and far-end signals degrades the \ac{AEC} performance~\cite{robledo-arnuncio_dealing_2007,pawig_adaptive_2010,abe_frequency_2014, ayrapetian_asynchronous_2021,helwani_clock_2022}. 

The degradation of \ac{AEC} in the presence of \ac{SRO} can be resolved through either asynchronous or synchronous methods. In asynchronous methods, the reference signals needed for \ac{AEC} is estimated by using, for example, a set of fixed beamformers at different non-overlapping directions followed by an adaptive reference adaptation algorithm~\cite{ayrapetian_asynchronous_2021}. This approach avoids the explicit need for synchronization. However, this method encounters a significant issue of near-end speech distortion when the near-end speech is captured by the beamformer. Furthermore, it relies on the assumption that the near-end speech activity is sparse, which holds true for certain applications like keyword spotting but may not necessarily apply to teleconferencing scenarios.

In synchronous methods, the far-end signal is initially synchronized with the microphone signal before running the \ac{AEC}. Pawig et al.~\cite{pawig_adaptive_2010} modeled the \ac{SRO} as a time scaling parameter. This parameter was then estimated and used for synchronizing the signals before applying the \ac{AEC}. Abe et al.~\cite{abe_frequency_2014} estimated the \ac{SRO} in the frequency domain using a simple extension of the least mean squares algorithm before rotating the phase of the far-end signal to approximate the time domain resampling. Helwani et al.~\cite{helwani_clock_2022} proposed a novel Kalman filtering approach which blindly accounts for the \ac{SRO}. However, these studies are primarily limited to single-device scenarios, assuming an \ac{SRO} between the microphone and loudspeaker due to mismatched sampling frequencies of the A/D and D/A converters.

In this study, we propose a synchronous method that addresses the two-device scenario as a two-channel \ac{AEC} problem with \ac{SRO} compensation. We assume that the primary device running the \ac{AEC} has access to all far-end signals, but only before they are transmitted to the auxiliary loudspeakers. \ac{SRO} compensation involves first estimating the \ac{SRO} between the devices and then resampling the far-end signal using the estimated \ac{SRO} before running the two-channel \ac{AEC}. For \ac{SRO} estimation, we use the robust \ac{DWACD} algorithm~\cite{gburrek_synchronization_2022} belonging to the family of coherence drift methods~\cite{schmalenstroeer_multi-stage_2017, chinaev_online_2021, gburrek_synchronization_2022}. We investigate the performance of both correlated and uncorrelated playback signals. Experiments in both echo-only and double-talk cases show that, for uncorrelated playback signals, it is possible to compensate for \ac{SRO}. However, for correlated playback signals, there is a performance gap between the proposed method with both oracle and estimated \acp{SRO} and the two-device AEC with no \ac{SRO}. Finally, we show that an independent single-channel \ac{AEC} filter is essential to ensure faster convergence of the estimated \ac{SRO}.   
\begin{figure}[t]
\includegraphics[width=0.9\linewidth]{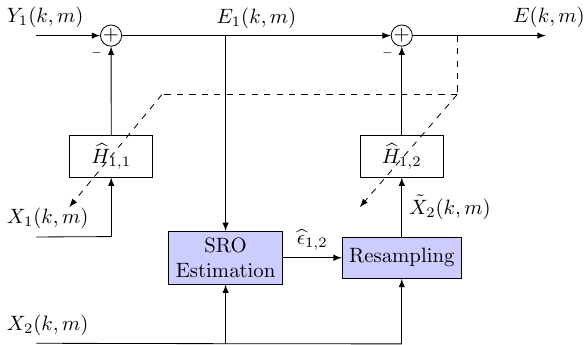}
\caption{Proposed two-device~\ac{AEC} envisioned as two-channel system with~\ac{SRO} compensation ("Variant 1").\vspace{-3pt}}
\label{fig:md_mc_system_v1}
\end{figure}
\vspace{-5mm}
\section{Multi-Device Scenario}
\label{sec:prob_formulation}
\vspace{-3mm}
Let us consider a room with $Q$ devices. Without loss of generality, we assume one microphone and one loudspeaker per device. The time-domain microphone signal of the primary device with index $p \in \{1,Q\}$ is given by:
\begin{equation}
\label{equ:mic_sig_SRO}
y_{p}\left[\frac{n}{f_{p}}\right] = \sum_{q=0}^{Q-1} h_{p,q}\left[\frac{n}{f_{p}}\right] * x_{q}\left[\frac{n}{f_{q}}\right] + z_{p}\left[\frac{n}{f_{p}}\right],
\end{equation}
where * represents the convolution, $n$ is the time index, $x_{q}[n/f_q]$ is the far-end signal played from the loudspeaker of the $q\,^\text{th}$ device, $h_{p,q}[n/f_p]$ is the~\ac{AIR} between the loudspeaker of the $q\,^\text{th}$ device and microphone of the $p^\text{th}$ device, $z_{p}[n/f_p]$ is the contribution of near-end talkers and noise at the microphone of the $p^\text{th}$ device. The sampling rate of $p^\text{th}$ and $q\,^\text{th}$ device are given by $f_{p}$ and $f_{q}$, respectively. 
The relation between $f_{p}$ and $f_{q}$ can be expressed in terms of \ac{SRO} between the devices $\epsilon_{p,q}$ as:
\begin{equation}
\label{equ:equ_SRO}
f_{q} = (1+\epsilon_{p,q})~f_{p}, 
\end{equation}
where $|\epsilon_{p,q}| \ll 1$ is usually expressed in~\ac{ppm}.

This study considers a two-device scenario with the first device ($p = 1$) as the primary device. We assume that i) there is no \ac{SRO} between the loudspeaker and microphone of the primary device ($\epsilon_{1,1} = 0$), and ii) there exists an unknown~\ac{SRO} $\epsilon_{1,2}$ between the loudspeaker and microphone signals not belonging to the same device ($|\epsilon_{1,2}| \geq 0$). With these assumptions, (\ref{equ:mic_sig_SRO}) becomes: 
\begin{equation}
\label{equ:mic_sig_SRO_v1}
y_{1}\left[\frac{n}{f_{1}}\right] = h_{1,1}[n]*x_{1}\left[\frac{n}{f_{1}}\right] + h_{1,2}[n]*x_{2}\left[\frac{n}{f_{2}}\right] + z_{1}\left[\frac{n}{f_{1}}\right].
\end{equation}  

Issues such as sampling time offset, variable \ac{SRO}, and packet loss during transmission are beyond the scope of this paper. Since the primary device, on which the \ac{AEC} is running, has no access to $x_{2}\left[\frac{n}{f_{2}}\right]$ in~\eqref{equ:mic_sig_SRO_v1}, we express this signal in terms of the sampling rate $f_1$ and \ac{SRO} $\epsilon_{1,2}$. For this purpose, substituting~\eqref{equ:equ_SRO} in~\eqref{equ:mic_sig_SRO_v1}, and applying Taylor series approximation to the term $x_{2}\left[\frac{n}{(1+\epsilon_{1,2})~f_{1}}\right]$~\cite{SMG_BSRO_2012},~\eqref{equ:mic_sig_SRO_v1} can be approximated in frequency domain with sufficiently long window size $N_\text{w}$ and hop size $N_\text{h}$ as:
\begin{multline}
\label{equ:mic_sig_SRO_modified_freqdomain_2device}
Y_{1}\left( k, m \right) = H_{1,1}\left(k,m\right)~X_{1}\left(k,m\right) + Z_{1}\left(k,m\right) \\
 + H_{1,2}\left(k,m\right)~\Lambda(k,m)~X_{2}\left(k,m\right), 
\end{multline}
where $\Lambda(k,m) = e^{\frac{-j2\pi k}{N_\text{w}}\left(\frac{mN_\text{h}\,\epsilon_{1,2}}{f_{1}}\right)}$. In~\eqref{equ:mic_sig_SRO_modified_freqdomain_2device}, $X_{1}\left(k,m\right)$ and $X_{2}\left(k,m\right)$ with frequency index $k$ and frame index $m$ are the frequency-domain representation of $x_{1}\left[\frac{n}{f_{1}}\right]$ and $x_{2}\left[\frac{n}{f_{1}}\right]$, respectively. $H_{p,q}(k,m)$ denotes the frequency-domain representation of $h_{p,q}[n/f_p]$. Note that \eqref{equ:mic_sig_SRO_modified_freqdomain_2device} holds only if the condition $\frac{mN_\text{h}\epsilon_{1,2}}{f_{1}} \ll N_\text{w}$ is satisfied~\cite{wang_CMSRO_WASN_2016}. 

The aim of two-device~\ac{AEC} running on the primary device is to eliminate the echoes from $Y_{1}\left( k, m \right)$ by modeling the~\ac{AIR}~$H_{p,q}(k,m)$ with the help of a linear~\ac{FIR} filters. However, when $X_{2}\left(k,m\right)$ is used as a reference signal, the linear~\ac{FIR} filter, which needs to model the \ac{AIR} $H_{1,2}(k,m)\,\Lambda(k,m)$, does not converge sufficiently due to the time varying term $\Lambda(k,m)$. To mitigate the convergence issue, we need to compensate for the term $\Lambda(k,m)$.
\vspace{-3mm}

\section{Proposed Method}
\label{sec:prop_system}
\begin{figure}[t]
\includegraphics[width=0.9\linewidth]{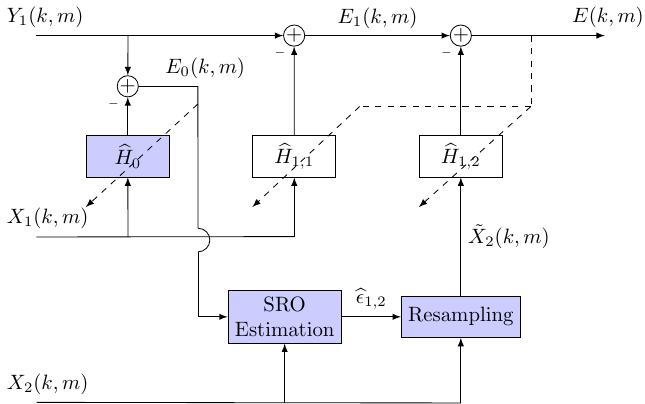}
\caption{Proposed two-device~\ac{AEC} envisioned as two-channel system with~\ac{SRO} compensation ("Variant 2").}
\label{fig:md_mc_system_v2}
\end{figure}

\vspace{-2mm}
The two variants of the proposed two-device~\ac{AEC} envisioned as a two-channel~\ac{AEC} system with~\ac{SRO} compensation are shown in Figs.~\ref{fig:md_mc_system_v1} and~\ref{fig:md_mc_system_v2}, respectively. The~\ac{SRO} compensation comprises of~\ac{SRO} estimation and resampling, which is described in Sec.~\ref{subsec:SRO_est}. The two-channel~\ac{AEC} filter is described in detail in Sec.~\ref{subsec:two_ch_Kalman}. The main difference between the two variants is the input signal used for \ac{SRO} estimation. Variant~1 uses the error signal $E_{1}$ as the input, while Variant~2 uses $E_{0}$. The error signal $E_{0}$ is calculated by subtracting the estimated echo from the microphone signal, where the estimated echo is obtained from an independent single-channel AEC filter $\widehat{H}_0$. For simplicity, in Sec.~\ref{subsec:SRO_est}, we refer the input signal to \ac{SRO} estimation as $I(k,m)$.

\begin{figure*}
\begin{subfigure}[b]{0.5\textwidth}
  \centering
  \includegraphics[width=\linewidth]{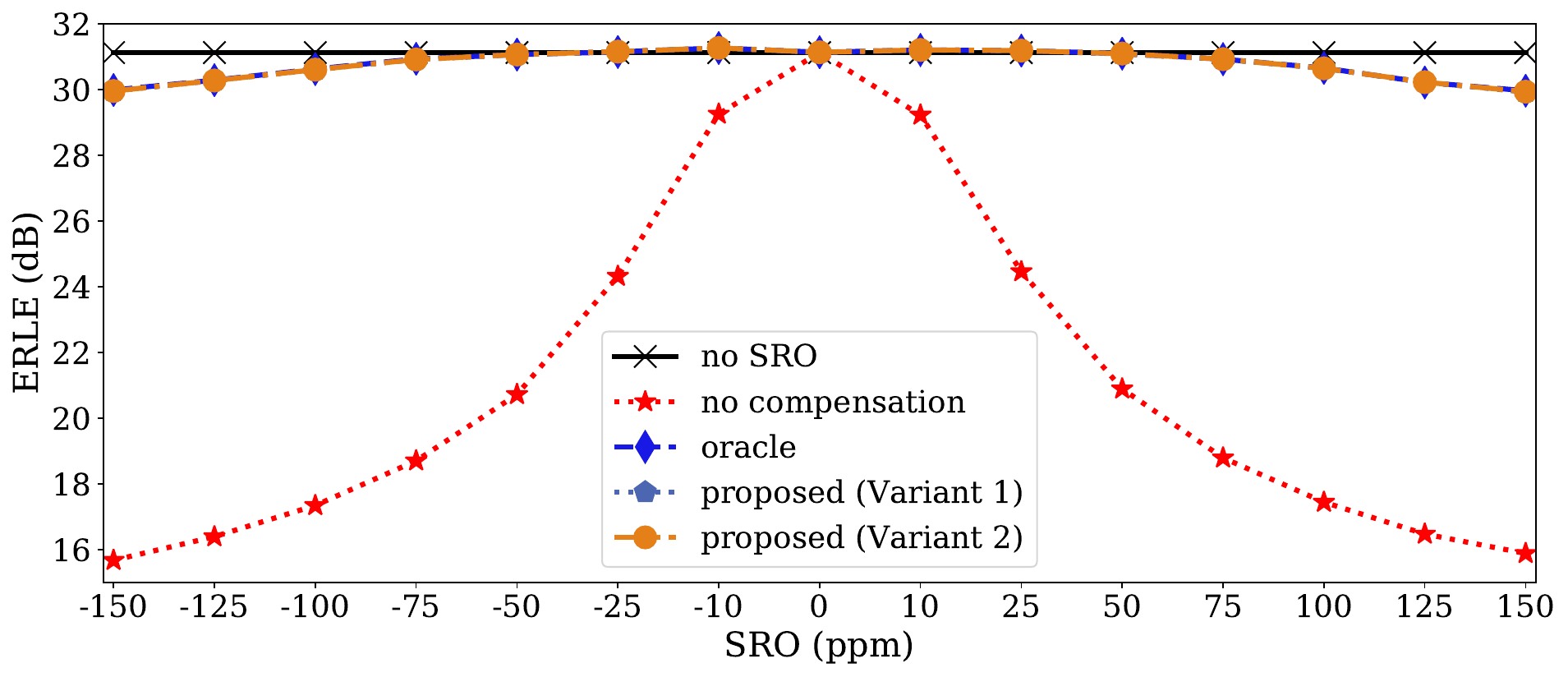}
  \label{fig:sfiga}
\end{subfigure}%
\begin{subfigure}[b]{0.5\textwidth}
  \centering
  \includegraphics[width=\linewidth]{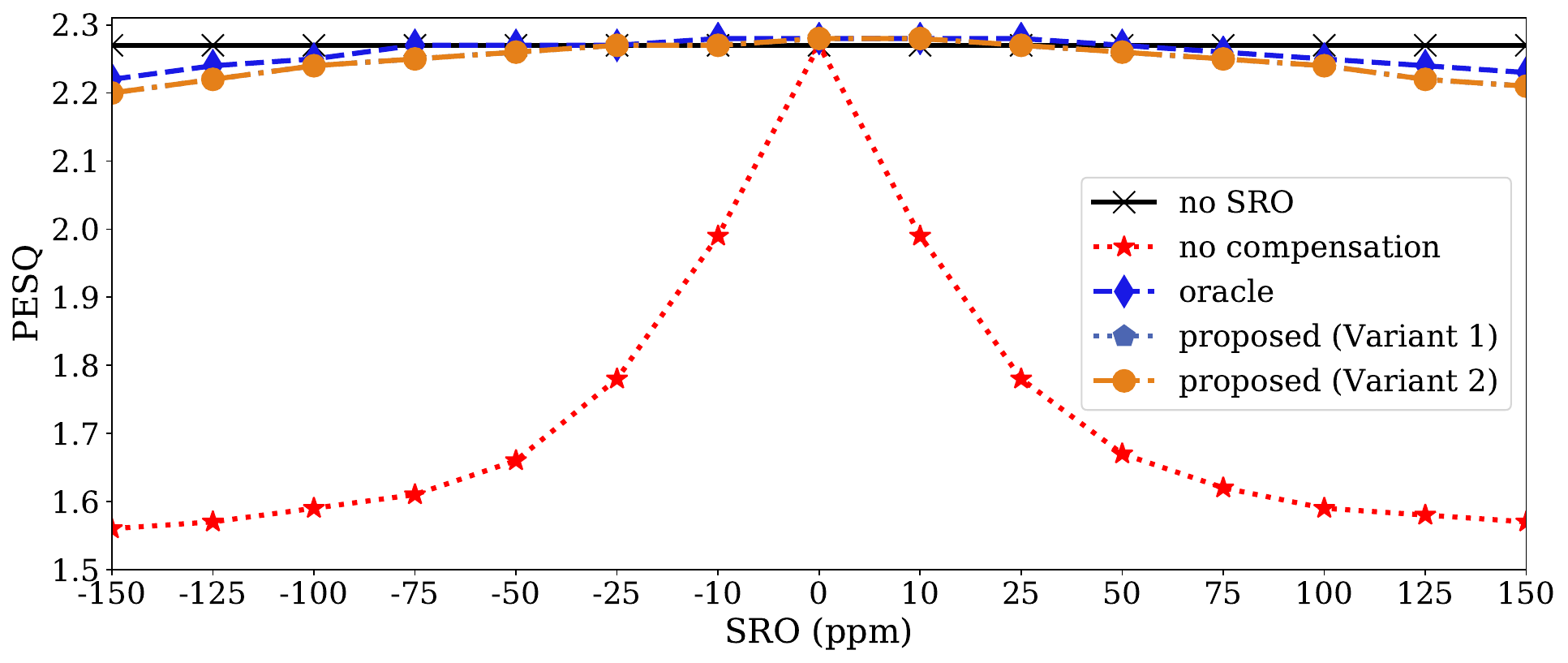}
  \label{fig:sfigb}
\end{subfigure}
\vspace{-10mm}
\caption{\Ac{ERLE}\,(\unit{dB}) vs.~\ac{SRO}\,(\,\unit{ppm}\,)\,(\,left\,) and PESQ vs.~\ac{SRO}\,(\,\unit{ppm}\,)(\,right\,) for uncorrelated playback signals.\vspace{-5pt} }
\label{fig:results_SROvsERLE_PESQ_uncorrelated}
\end{figure*} 

\vspace{-3mm}
\subsection{SRO compensation}
\label{subsec:SRO_est}
\vspace{-1.5mm}
To estimate the~\ac{SRO}, we use the~\ac{DWACD} algorithm proposed in~\cite{gburrek_synchronization_2022}. Given the input signal $I(k,m)$ and the reference signal $X_{2}(k,m)$, the \ac{SRO} estimation relies on first computing the complex coherence function $\Gamma(k,m)$, which is given by:
\begin{equation}
\label{coherence_func_ACD}
\Gamma(k,m) = \frac{\Phi_{I~X_{2}}(k,m)}{\sqrt{\Phi_{II}(k,m)\Phi_{X_{2}~X_{2}}(k,m)}}, 
\end{equation}
where $\Phi_{I~X_{2}}$, $\Phi_{II}$ and $\Phi_{X_{2}~X_{2}}$ are the cross and auto power spectral densities, respectively. The phase function $\tilde{P}(k,m)$ is then computed by the complex conjugated product of two consecutive complex coherence functions with a temporal distance of $L$. It is given by: 
\begin{equation}
\label{avg_coherence_func_wACD}
\tilde{P}(k,m) = \Gamma(k,m+L)~\Gamma^*(k,m),
\end{equation}
where $^*$ denotes the complex conjugate. The temporally averaged phase function $P(k,m)$ and the \ac{GCC} $p(\beta,m)$ are given by:
\begin{equation}
\label{avg_coherence_func_dwACD}
P(k,m) = \alpha~P(k,m-1) + (1-\alpha)~\tilde{P}(k,m),
\end{equation}
\begin{equation}
\label{equ:gcc}
p(\beta,m) = \textrm{IDFT}\{P(k,m)\},
\end{equation}
where $\alpha$ is the smoothing factor, $\beta$ is the time-lag and $\textrm{IDFT}$ is the inverse DFT. 
The estimated~\ac{SRO} $\widehat{\epsilon}_{1,2}$ is obtained by first finding the integer time-lag $\beta_\text{max}$ that maximizes the~\ac{GCC}: 
\begin{equation}
\label{SRO_estimate_dwACD_1}
\widehat{\epsilon}_{1,2}(m) = -\frac{1}{L\,N_{\text{h}}}\beta_\text{max} = -\frac{1}{L\,N_{\text{h}}}\,\arg \max_{\beta}|p(\beta,m)|.
\end{equation} 
Then, an accurate \ac{SRO} estimate is obtained by determining the non-integer time-lag by performing a golden search in the interval given by $[\beta_\text{max}-0.5,\beta_\text{max}+0.5]$. 
The complex coherence function $\Gamma(k,m)$ is estimated only when speech activity is detected in both the signals $X_{2}(k,m)$ and $I(k,m)$. In this study, we used the energy-based ideal VAD to detect speech activity.

Once $\widehat{\epsilon}_{1,2}$ is estimated, we can compensate for the the~\ac{SRO}. For this purpose, the term $\Lambda(k,m)$ is computed using $\widehat{\epsilon}_{1,2}$. Afterwards, we can obtain the resampled reference signal $\tilde{X}_{2}\left(k,m\right) = X_{2}\left(k,m\right)\,\Lambda(k,m)$ to be used in the \ac{AEC}. In addition, buffer management in both \ac{SRO} estimation and resampling is essential to prevent the accumulation of sample drifts~\cite{gburrek_synchronization_2022}. 

\vspace{-3mm}
\subsection{Two-channel~\ac{AEC} filter}
\label{subsec:two_ch_Kalman}
\vspace{-1.5mm}
The two-channel~\ac{AEC} filter is implemented as partitioned block frequency domain Kalman filter based on state-space architecture~\cite{kuech_state-space_2014}. The filters and each partition within the filters are assumed to be mutually uncorrelated with zero mean. Under such assumption and the assumptions made by Kuech et al. in~\cite{kuech_state-space_2014}, the $b^\text{th}$ partition of filter $\mathbf{\widehat{H}}_{p,q}$ and Kalman gain $\mathbf{K}^{b}_{p,q}(m)$ is given by:
\begin{equation}
\label{equ:two_ch_KF_FE}
\mathbf{\widehat{H}}^{b}_{p,q}(m + 1) = A\,\left[\,\mathbf{\widehat{H}}^{b}_{p,q}(m) + \mathbf{K}^{b}_{p,q}(m)\,\mathbf{E}(m)\,\right],
\end{equation}
\begin{multline}
\label{equ:two_ch_KF_KG}
\mathbf{K}^{b}_{p,q}(m) = \mathbf{P}^{b}_{p,q}(m)\left(\mathbf{X}^{b}_{q}\right)^\text{H}(m) \\
\left[ \sum_{b=0}^{B-1} \mathbf{X}^{b}_{q}(m-1) \mathbf{P}^{b}_{p,q}(m) \left(\mathbf{X}^{b}_{q}\right)^\text{H}(m)  + \frac{M}{V}~\Psi_{ZZ}(m) \right]^{-1},
\end{multline}
where $\mathbf{\widehat{H}}^{b}_{p,q}(m) = \left[ \widehat{H}^{b}_{p,q}(0,m), \widehat{H}^{b}_{p,q}(1,m), \ldots , \widehat{H}^{b}_{p,q}(k,m) \right]^{T}$, 
A is the transition factor, $T$ is the transpose, $\mathbf{E}(m)$ is the error signal, superscript H denotes the conjugate transpose of a matrix, $\Psi_{ZZ}(m)$ is the covariance of the near-end spectrum $Z(m)$, $\mathbf{X}^{b}_{q}$ is the $q^{th}$ far-end signal, $M$ and $V$ are the hop size and frame length respectively.
The covariance matrix of the coefficient error vector $\mathbf{P}^{b}_{p,q}(m)$ is given by: 
\begin{multline}
\label{equ:two_ch_KF_COVMAT}
\mathbf{P}^{b}_{p,q}(m) = A^2\left[\mathbf{I}_M - \mathbf{K}^{b}_{p,q}(m-1)\mathbf{X}^{b}_{q}(m-1)\right]\mathbf{P}^{b}_{p,q}(m-1) \\
 + \Psi_{b,\Delta\Delta}(m), 
\end{multline}
where $\Psi_{b,\Delta\Delta}$ is the covariance of the temporal variations of the acoustic transfer path. To ensure the convergence of the linear \ac{FIR} filter $\mathbf{\widehat{H}}^{b}_{1,2}$, we use $\tilde{X}_{2}\left(k,m\right)$ instead $X_{2}\left(k,m\right)$ of  as the reference signal.

\vspace{-2mm}
\section{Experimental Results}
\label{sec:exp_Res}
\vspace{-2mm}
The performance of our proposed system is evaluated under both correlated cases, where the devices are playing back the same signal, and uncorrelated cases, where the devices are playing back different signals. In the echo-only scenario,  \acf{ERLE}~\cite{Haensler_2004_book} is used to evaluate the performance whereas, \ac{PESQ}~\cite{itu_t_pesq} is used to assess the performance in the double-talk scenario by comparing the output of the proposed system with that of ground-truth near-end speech. For the evaluation, the above-mentioned metrics were computed by averaging the result over 50 test files.


\vspace{-4mm}
\subsection{Test signal generation} 
\vspace{-2mm}
To generate the microphone signals, we first created far-end and near-end signals of duration 36\,\unit{s} at a sampling rate of 16\,\unit{kHz} from the training set $si\_tr\_s$ of WSJ0 database ~\cite{wsj0_dataset_2016}. These files were normalized to -25\,\unit{dBFS} before convolving with simulated~\acp{RIR}~\cite{habets2008room}. The size of the room was chosen randomly with a uniform distribution between [5,5,3]\,\unit{m} and [8,8,6]\,\unit{m}. The reverberation time was set to a random value between 0.2\,\unit{s} and 0.5\,\unit{s} with uniform probability. The device positions and near-end talker positions were randomly chosen within the room. However, it was ensured that they were at least 0.75m away from the walls. Constant \acp{SRO} in the range of $\pm 150$\,\unit{ppm}~\cite{SchGbHaeb2023} were simulated on the echo signals using the STFT method proposed in~\cite{Schmalenstroeer_overlapsave_resampling_eusipco_2018} using segment length of 8192 samples~\cite{gburrek_synchronization_2022}. All the \ac{SRO} simulated echo signals were added along with near-end speech, if present, before adding a Gaussian noise at 40\,\unit{dB} SNR to simulate sensor noise. 

\vspace{-4mm}
\subsection{Evaluation}
\vspace{-2mm}
We compare the two variants of our system (Variant~1 and Variant~2) against systems assuming no \ac{SRO} between the devices (no SRO), no \ac{SRO} compensation (no compensation), and oracle \ac{SRO} compensation (oracle)\footnote{\url{https://www.audiolabs-erlangen.de/resources/2024-IWAENC-SRO-MDAEC}}. For the echo-only scenario, \ac{ERLE} is computed using the last 30\,\unit{s}, whereas, for the double-talk scenario, \ac{PESQ} is calculated using the entire signal. For the estimation of the \ac{SRO}, two previous segments, each of length 8192 samples corresponding to 0.512\,\unit{s}, are used. Hence, the estimation of the \ac{SRO} does not add any additional delay. The power spectral densities are computed by the Welch method with a hop length of 512 samples (32\,\unit{ms}), and the smoothing parameter $\alpha$ is set to 0.95. The Kalman filters are implemented with 10 taps, an FFT length of 512 samples, and a hop length of 256 samples, which corresponds to a filter length of 160\,\unit{ms} in the time domain. The transition parameter (A) is set to 0.999~\cite{kuech_state-space_2014}. 

Figure \ref{fig:results_SROvsERLE_PESQ_uncorrelated} evaluates the performance of the compared systems for echo-only and double-talk scenarios for uncorrelated playback signals. When the \ac{SRO} is uncompensated, the performance of the system drops since \ac{AEC} filter $\widehat{H}_{1,2}$ fails to converge to the true \ac{AIR} $H_{1,2}$. When compensated for the \ac{SRO}, both the variants of the proposed system reach the performance of the system that assumes no \ac{SRO} between the devices, especially at \ac{SRO} values smaller than $\pm$\,75\,\unit{ppm}. In addition, the performance of these variants matches that of the oracle system. The performance in the double-talk scenario is similar to the echo-only scenario except for a minor gap with the system that uses oracle \ac{SRO}, especially at higher \ac{SRO} values. 

\begin{figure}[t]
\begin{subfigure}[b]{0.5\textwidth}
  \centering
  \includegraphics[width=\linewidth]{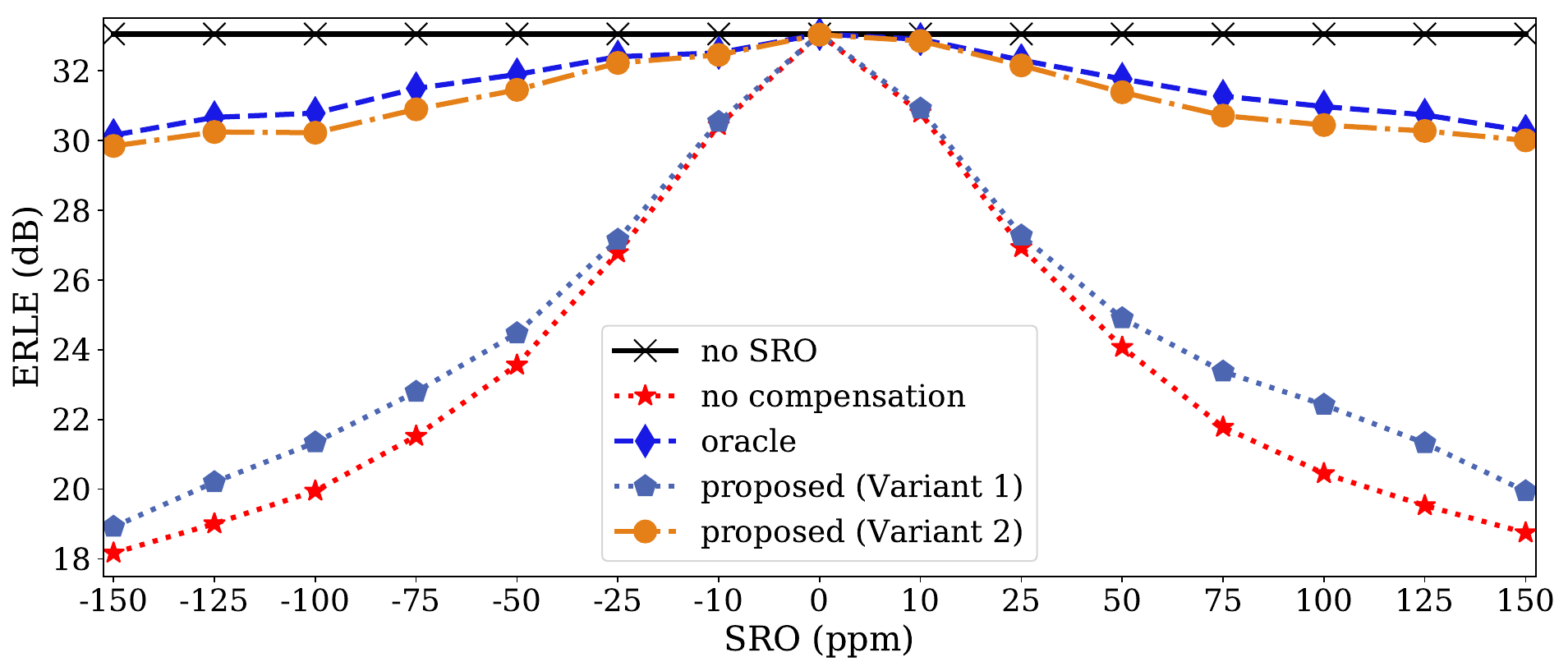}
\end{subfigure}
\begin{subfigure}[b]{0.5\textwidth}
  \centering
  \includegraphics[width=\linewidth]{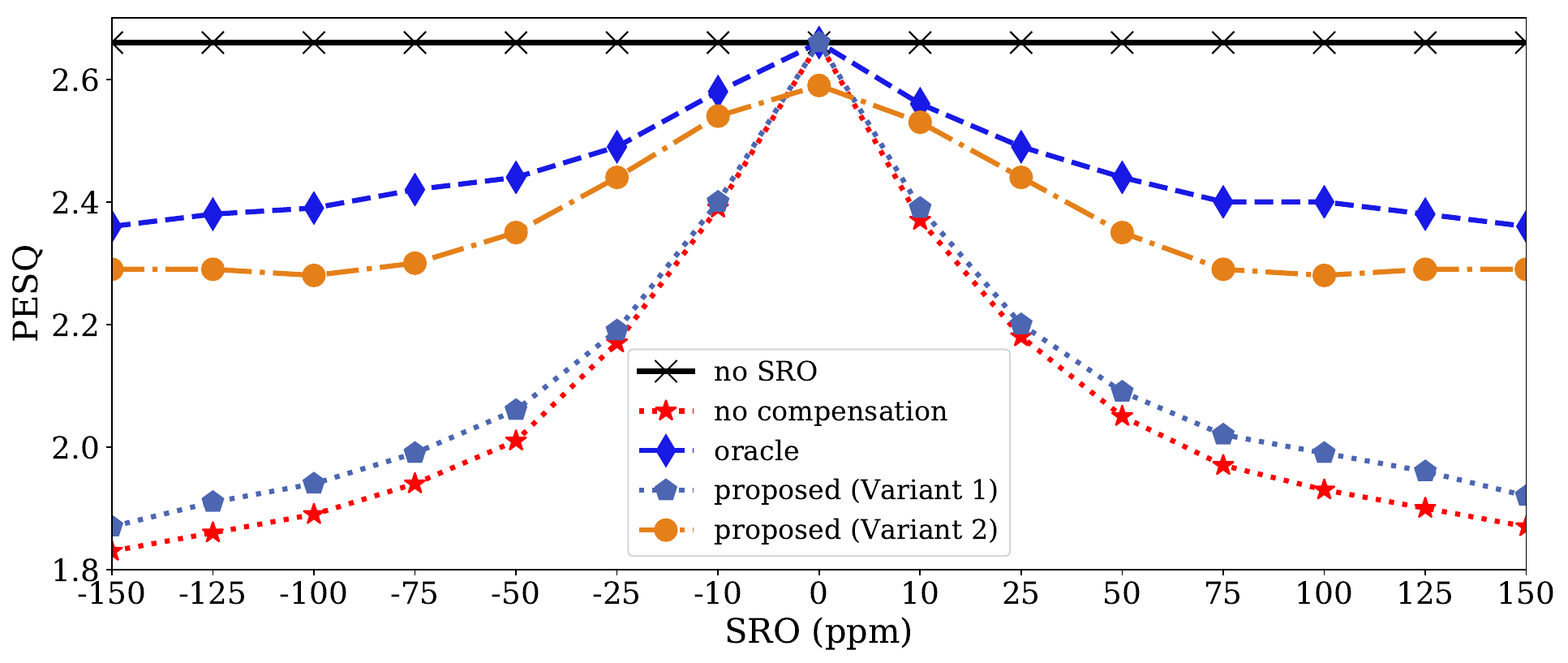}
\end{subfigure}
\caption{Top: \Ac{ERLE}\,(\unit{dB}) vs. \ac{SRO}(\,\unit{ppm}\,), Bottom: \ac{PESQ} vs. \ac{SRO}\,(\,\unit{ppm}\,) for correlated playback signals.}
\label{fig:results_SROvsPESQ_Correlated}
\end{figure}
Figure \ref{fig:results_SROvsPESQ_Correlated} evaluates the performance of the compared systems for echo-only and double-talk scenarios for correlated playback signals. When the \ac{SRO} is uncompensated, the system's performance drops due to the divergence of the two-channel \ac{AEC}. In Variant 1 of the proposed system, the filter convergence and \ac{SRO} estimation impact each other, leading to slow convergence of \ac{SRO}. Hence, the performance of Variant 1 is similar to that of the system with no compensation in both echo-only and double-talk scenarios. Hence, in Variant 2 of the proposed system, an independent single-channel \ac{AEC} filter $\widehat{H}_{0}$ is employed to decouple the filter convergence and \ac{SRO} estimation. This decoupling helps the estimated \ac{SRO} to converge faster to the true \ac{SRO}. In the echo-only scenario, the performance of Variant 2 of the proposed system and the system using oracle \ac{SRO} has almost identical performance. However, in the double-talk scenario, the performance of variant 2 of the proposed system fails to match the system's performance using oracle \ac{SRO}. The main reason for this is the inaccuracies in the estimated \ac{SRO} since the \ac{SRO} estimation in the double-talk scenario is not as robust as in the echo-only scenario. Variant 2 of the proposed system and the oracle system fail to compensate for the effect of \ac{SRO}. This is because, although the \ac{SRO} introduces some decorrelation, the filter $\widehat{H}_{1,1}$ and $\widehat{H}_{1,2}$ do not converge to the true \ac{AIR} $H_{1,1}$ and $H_{2,2}$ respectively.  

\vspace{-3mm}
\subsection{SRO estimation in the presence of echo path change}
\vspace{-2mm}
\begin{figure}[t]
\centering
  \includegraphics[width=0.95\linewidth]{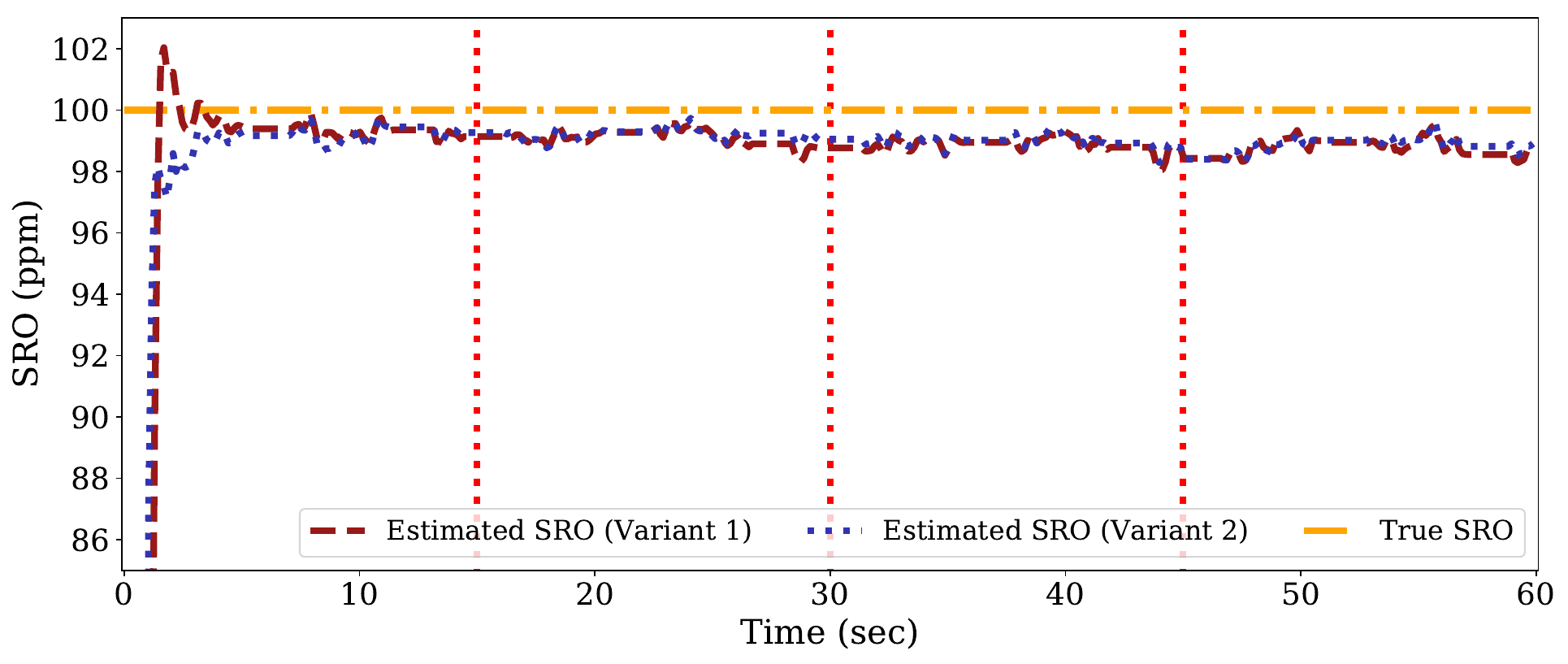}
  \caption{ Estimated \ac{SRO} (\,\unit{ppm}\,) vs. Time (\,\unit{sec}\,) during echo-only scenario with echo path changes for uncorrelated playback signals. True~\ac{SRO} is 100~\unit{ppm}. }
  \label{fig:ofe_epc_uncorr}
\end{figure}
\begin{figure}[t]
\centering
  \includegraphics[width=0.95\linewidth]{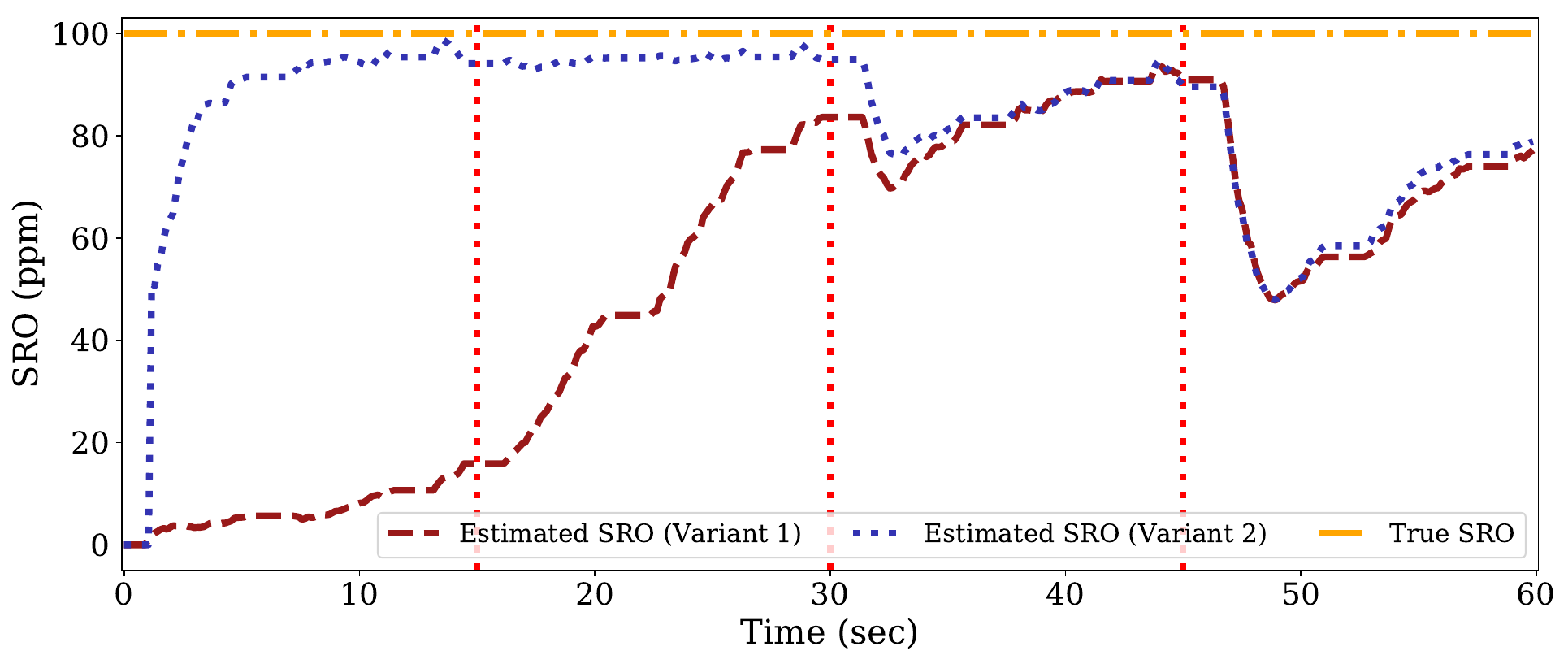}
  \caption{ Estimated \ac{SRO} (\,\unit{ppm}\,) vs. Time (\,\unit{sec}\,) during echo-only scenario with echo path changes for correlated playback signals. True~\ac{SRO} is 100~\unit{ppm}.}
  \label{fig:ofe_epc_corr}
\end{figure}

In this section, we evaluate the robustness of \ac{SRO} estimation in echo-only scenarios for uncorrelated and correlated playback signals. We simulate three echo path changes on a signal with duration 60\,\unit{s}. At 15\,\unit{s}, only the position of device two is changed, thus affecting the \ac{AIR} $H_{1,2}$. At 30\,\unit{s}, only the microphone of device one is changed, affecting the \ac{AIR} $H_{1,1}$. At 45\,\unit{s}, both the microphone of device one and device two are simultaneously changed, affecting the \ac{AIR} $H_{1,2}$ and $H_{1,1}$ simultaneously. 

Figure \ref{fig:ofe_epc_uncorr} compares the estimated \ac{SRO} for the two variants of the proposed system with the true \ac{SRO} for uncorrelated playback signals. The estimated \ac{SRO} for both the variants quickly converges to the true \ac{SRO} and is robust to all the simulated echo path changes. 

Figure \ref{fig:ofe_epc_corr} compares the estimated \ac{SRO} for the two variants of the proposed system with the true \ac{SRO} for correlated playback signals. The \ac{SRO} estimation converges to true \ac{SRO} faster in Variant 2 when compared to Variant 1. Once the \ac{SRO} estimation converges, \ac{SRO} estimation of both system variants performs similarly. In addition, we can observe that change in \ac{AIR} $H_{1,2}$ has no impact on the \ac{SRO} estimation, whereas change in \ac{AIR} $H_{1,1}$ leads to a small drop in the estimated \ac{SRO}. However, a simultaneous change in \ac{AIR} $H_{1,2}$ and $H_{1,1}$ simultaneously shows a huge drop in the estimated \ac{SRO}. 

\vspace{-3mm}
\section{Conclusion}
\label{sec:conlusion}
\vspace{-3mm}
We proposed two variants of two-channel AEC for addressing the two-device AEC problem in the presence of \ac{SRO} and evaluated them for both uncorrelated or correlated playback signals in echo-only and double-talk scenarios. We showed that for uncorrelated playback signals, it is possible to compensate for \ac{SRO} and reach the performance of the system with no \ac{SRO}. Additionally, we demonstrated that the \ac{SRO} estimation is robust to echo path changes. For the correlated playback signals, to ensure faster convergence of \ac{SRO}, we use an independent single-channel \ac{AEC} filter. This is primarily done to decouple the filter convergence from the \ac{SRO} estimation. 

\flushend
\bibliographystyle{IEEEbib}
\bibliography{refs}

\end{document}